\documentclass[twocolumn,showpacs,amsmath,amssymb]{revtex4}
\usepackage{graphicx}% Include figure files
\usepackage{bm}% bold math

\begin{document}

%\preprint{APS/123-QED}

\title{A numerical method of solving time-dependent Hartree-Fock-Bogoliubov 
equation with Gogny interaction}% Force line breaks with \\

\author{Y.~Hashimoto and K.~Nodeki}
\affiliation{%
Graduate School of Pure and Applied Sciences, 
University of Tsukuba, Tsukuba 305-8571, Japan 
}%

\date{\today}% It is always \today, today,
             %  but any date may be explicitly specified

\begin{abstract}
A numerical method to solve time-dependent Hartree-Fock-Bogoliubov equation 
is proposed to treat the case with Gogny effective interaction. 
To check the feasibility of the method, it is applied to oxygen isotope ${}^{20}$O  
and small amplitude oscillations are calculated. The conservation of the nucleon 
numbers as well as energy expectation value is demonstrated. 
The strength distributions of the small amplitude quadrupole 
oscillations are also shown.
\end{abstract}

% PACS, the Physics and Astronomy
                             % Classification Scheme.
%\keywords{Suggested keywords}%Use showkeys class option if keyword
                              %display desired
\maketitle

\section{Introduction}
The structure and dynamics of exotic nuclei have been 
the main subject of investigation 
both in the theoretical and the experimental nuclear physics.
The peculiar feature of the exotic nuclei is that 
the Fermi level is located near the continuum levels. 
In this situation, the nucleons near the Fermi level 
are easily brought to the continuum states by pairing 
correlations. The low-energy excitation modes are often 
in the continuum energy region.

The general mean-field method to treat the pairing correlations 
as well as the mean-field is the Hartree-Fock Bogoliubov (HFB) method, 
which has played a central role in the investigation of the 
static ground state properties of the nuclei 
in a wide area of the nuclear chart \cite{RevModPhys}. 
The typical method to study the excited collective states of 
nuclei on top of the mean-field 
ground state with the pairing correlations 
is the quasiparticle random phase approximation (QRPA) 
\cite{engel,colo,bender,matsuo,khan,giambrone,paar,goriely,terasaki}. 

In the practical QRPA calculations, two major methods so far developed are  
the Green's function method through the response function formalism 
and the diagonalization method of the QRPA matrix. 
From the theoretical point of view, 
the QRPA is a small amplitude approximation of the 
time-dependent Hartree-Fock Bogoliubov (TDHFB) equation \cite{RS}.
Considering the rich results of the time-dependent Hartree-Fock 
(TDHF) and the random phase approximation (RPA) 
as a small amplitude approximation of the TDHF \cite{negele}, 
it is worthwhile establishing a self-consistent TDHFB framework 
together with the practical method of solving the TDHFB equation. 

The widely used effective interactions both in the mean-field 
and the QRPA calculations are the Skyrme interactions  
with zero-range pairing part or the Gogny interaction   
which consists of both the finite-range parts and  
the zero-range ones.

In the case of the HFB calculations with the Skyrme interactions (Skyrme HFB), 
the Skyrme interaction is used for the particle-hole channel,
while the pairing interaction is introduced 
only for the particle-particle channel.
Since the zero-range interaction is assumed in the Skyrme HFB, 
it is necessary to set the appropriate cut-off energy and choose 
the optimum parameter set in the pairing part \cite{Borycki}. 

In contrast with the Skyrme HFB, in the HFB calculations 
with the Gogny interaction (Gogny HFB), the particle-hole channel and 
the particle-particle channel are treated on an equal footing.
Therefore, the Gogny interaction is suitable for the formulation 
of the self-consistent TDHFB together with a practical numerical 
method of integrating the TDHFB equation.

In this paper, we report the formulation of the 
self-consistent TDHFB with the Gogny interaction and 
a numerical method to integrate the TDHFB equations.
With the aim at illustrating the feasibility of the method,  
we apply the method to the case of oxygen isotope ${}^{20}$O, 
carrying out the numerical integration of the TDHFB equations.   

This paper consists of the following sections: In Sec. II, a derivation  
of the self-consistent TDHFB equation is given  
together with a numerical method of solving the TDHFB equation. 
In Sec. III, the results of applying the TDHFB equation 
to the oxygen isotope ${}^{20}$O are shown.
Section IV is for summary and concluding remarks.

\section{TDHFB equation and numerical solution}
\subsubsection{TDHFB equation}
Let us assume that the Hamiltonian is written as a sum of the kinetic energy 
and the two-particle interactions with particle creation (annihilation) operator 
$C_{\alpha}^{\dagger}$ ($C_{\alpha}$),
\begin{eqnarray}
  H = \sum_{\alpha \beta} T_{\alpha \beta} C_{\alpha}^{\dagger} C_{\beta} 
     + \frac{1}{4} \sum_{\alpha \beta \gamma \delta} 
         {\cal V}_{\alpha \beta \gamma \delta} C_{\alpha}^{\dagger} C_{\beta}^{\dagger}  
                     C_{\delta} C_{\gamma}, \label{origH} 
\end{eqnarray}
where $T_{\alpha \beta}$ is the kinetic energy matrix element and 
${\cal V}_{\alpha \beta \gamma \delta}$ is an antisymmetrized two-body matrix element.

In the formal presentation of the TDHFB equation, it is convenient 
to start with the generalized density matrix ${\cal R}$ \cite{RS},
\begin{eqnarray} 
  {\cal R} = \left( 
             \begin{array}{cc}
               \rho & \kappa \cr
              - \kappa^{*} & 1 - \rho^{*} 
             \end{array}             
             \right)\, , \label{GdenstMat}
\end{eqnarray}
where $\rho$ and $\kappa$ are normal density and pairing tensor, respectively, 
\begin{eqnarray}
 \rho_{\alpha \beta} &=& \left( V^{*} V^{T} \right)_{\alpha \beta},  \label{rhoVV} \\
 \kappa_{\alpha \beta} &=& \left( V^{*} U^{T} \right)_{\alpha \beta}. \label{kapVU} 
\end{eqnarray}
The matrices $U_{\alpha k}$ and $V_{\alpha k}$ are introduced to 
connect the particle operators \{$C_{\alpha}^{\dagger}$, $C_{\alpha}$\} with 
the quasiparticle operators \{$\beta_k^{\dagger}$, $\beta_k$\} as follows:
\begin{eqnarray}
   \beta_{k}^{\dagger} &=& 
    \sum_{\alpha} \left( U_{\alpha k} C_{\alpha}^{\dagger}  
                       + V_{\alpha k} C_{\alpha} \right), \label{Bog-1} \\
   \beta_{k}           &=& 
    \sum_{\alpha} \left( U_{\alpha k}^{*} C_{\alpha}
                       + V_{\alpha k}^{*} C_{\alpha}^{\dagger}   \right). \label{Bog-2}
\end{eqnarray}

The equation of motion of the generalized density matrix 
${\cal R}$ in Eq. (\ref{GdenstMat}) is given by
\begin{eqnarray}
   i \hbar \dot{{\cal R}} = \left[ {\cal H}, {\cal R} \right], \label{eq_motion}
\end{eqnarray}
with the HFB Hamiltonian ${\cal H}$ given by
\begin{eqnarray}
  {\cal H} = \left( 
             \begin{array}{cc}
               h & \Delta \cr
              - \Delta^{*} & - h^{*} 
             \end{array}             
             \right)\, .  \label{hfbH}
\end{eqnarray}
Here, mean field Hamiltonian $h$ and pairing mean field $\Delta$ are 
introduced through the following relations:
\begin{eqnarray}
  h_{\alpha \beta} &=& T_{\alpha \beta} + \Gamma_{\alpha \beta}, \quad 
 \Gamma_{\alpha \beta} 
  = \sum_{\gamma \delta} {\cal V}_{\alpha \gamma \beta \delta} \rho_{\delta \gamma}, \label{GamDel-1} \\
 \Delta_{\alpha \beta} &=& \frac{1}{2}\sum_{\gamma \delta} {\cal V}_{\alpha \beta \gamma \delta} \kappa_{\gamma \delta}.
     \label{GamDel-2}
\end{eqnarray}
  
From the definitions in Eqs. (\ref{rhoVV}) and (\ref{kapVU}),  
the generalized density matrix ${\cal R}$ in Eq. (\ref{GdenstMat}) is rewritten as
\begin{eqnarray}
  {\cal R} = \left(
             \begin{array}{c}
                V^{*} \cr
                U^{*}
             \end{array}
             \right)
             \left( V^{T}, U^{T} \right), \label{rewritten}
\end{eqnarray}
where $T$ stands for transposed matrix.
Combining the formal solution ${\cal R}(t)$ of the equation of motion 
in Eq. (\ref{eq_motion}) given as, 
\begin{eqnarray}
  {\cal R}(t) = {\rm e}^{- \frac{i}{\hbar} \int^{t} d \tau {\cal H}(\tau)} {\cal R}(0) 
                {\rm e}^{  \frac{i}{\hbar} \int^{t} d \tau {\cal H}(\tau)},  
\end{eqnarray}
with the rewritten form of the generalized density matrix ${\cal R}$ 
in Eq. (\ref{rewritten}), we have a formal solution of the matrices $U$ and $V$ 
as follows:
\begin{eqnarray}
   \left(
    \begin{array}{c}
                V^{*}(t) \cr
                U^{*}(t)
             \end{array}
             \right)
   = {\rm e}^{- \frac{i}{\hbar} \int^{t} d \tau {\cal H}(\tau)} 
   \left(
    \begin{array}{c}
                V^{*}(0) \cr
                U^{*}(0)
             \end{array}
             \right). \label{formalUV}
\end{eqnarray}
The formal solution Eq. (\ref{formalUV}) of the matrices $U$ and $V$ is equivalent to the 
equation of motion of the matrices $U$ and $V$ given as
\begin{eqnarray}
  i \hbar \frac{\partial}{\partial t} 
    \left(
    \begin{array}{c}
                V^{*}(t) \cr
                U^{*}(t)
             \end{array}
             \right)
     &=& {\cal H} \left( 
    \begin{array}{c}
                V^{*}(t) \cr
                U^{*}(t)
             \end{array}
             \right),                       \nonumber \\  
  i \hbar \frac{\partial}{\partial t} 
    \left(
    \begin{array}{c}
                U(t) \cr
                V(t)
             \end{array}
             \right)
     &=& {\cal H} \left( 
    \begin{array}{c}
                U(t) \cr
                V(t)
             \end{array}
             \right), \label{hfbeq_UV}
\end{eqnarray}
where the definition of the HFB Hamiltonian ${\cal H}$ in Eq. (\ref{hfbH}) 
is used. This form of the TDHFB equation was used by Bulgac in 
relation with Berry's phase \cite{Bul}.

\subsubsection{Numerical method of solution}
The TDHFB equation (\ref{hfbeq_UV}) takes a simple form, being similar to 
the TDHF equation with a TDHF Hamiltonian $h_{TDHF}$,
\begin{eqnarray}
   i \hbar \frac{\partial}{\partial t} \psi_j({\bf x}, t) 
     = h_{TDHF} \psi_j({\bf x}, t), \label{tdhfeq}
\end{eqnarray}
for the wave functions $\psi_j({\bf x}, t) (j = 1, 2, \cdots, {\rm N})$ 
of N orbitals.
The solution of the TDHF equation is calculated by making use of the 
relation,
\begin{eqnarray}
 \psi_j({\bf x}, t_{n+1}) = {\rm e}^{-i \frac{\Delta t}{\hbar} h^{(n+1/2)}} \psi({\bf x}, t_{n}), 
     \label{exp-h}
\end{eqnarray}
at every time step from $t_{n}$ to $t_{n+1} = t_{n} + \Delta t$ 
with an adequate Hamiltonian $h^{(n+1/2)}$ to conserve the total energy \cite{Floc}.
Applying the method in (\ref{exp-h}) to the TDHFB equation (\ref{hfbeq_UV}), 
we get the solution of the TDHFB equation (\ref{hfbeq_UV}) in the form given as
\begin{eqnarray}
        \left(
        \begin{array}{c}
                U \cr
                V
             \end{array}
             \right)^{(n+1)} = \exp \left( -i \frac{\Delta t}{\hbar} {\cal H}^{(n+1/2)} \right) 
    \left(
    \begin{array}{c}
                U \cr
                V
             \end{array}
             \right)^{(n)}, \label{tdhfb-solution}
\end{eqnarray} 
with an adequate TDHFB Hamiltonan ${\cal H}^{(n+1/2)}$ 
at every time step from $t_{n}$ to $t_{n+1} = t_{n} + \Delta t$.

In the present case with the Hamiltonian (\ref{origH}), the expectation value 
of the Hamiltonian (\ref{origH}) with respect to HFB state $|\Phi \rangle$ is 
given as
\begin{eqnarray}
  E &=& \langle \Phi | H | \Phi \rangle \nonumber \\ 
    &=& \sum_{\alpha \beta} T_{\alpha \beta} \rho_{\beta \alpha} 
   + \frac{1}{2} \Gamma_{\alpha \beta} \rho_{\beta \alpha} 
   + \frac{1}{2} \kappa_{\alpha \beta}^{*} \Delta_{\alpha \beta}, \label{Ehfb}
\end{eqnarray} 
where the mean potential $\Gamma$ and mean pairing potential $\Delta$ are defined 
in Eqs. (\ref{GamDel-1}) and (\ref{GamDel-2}). 
Since in the TDHFB calculation the energy conservation is one 
of the most important conditions to be fulfilled, let us see how the 
energy in (\ref{Ehfb}) is conserved with respect to a small variation 
in the matrices $U$ and $V$. 
Within the first order of the parameter $\lambda \equiv \frac{\Delta t}{\hbar}$ 
the matrices $U$ and $V$ at a time $t$ is changed into the new matrices $U'$ and $V'$
according to the following relation,
\begin{eqnarray}
 \left( \begin{array}{c}
         U' \\
         V'  \end{array}
 \right)
&=&  \left( \begin{array}{c}
         U \\
         V  \end{array}
 \right)
 - i \lambda \left( \begin{array}{cc}
                    h & \Delta \\
                  - \Delta^{*} & - h^{*} 
                   \end{array}
             \right) 
      \left( \begin{array}{c}
         U \\
         V  \end{array}
      \right) \nonumber \\
 &=& \left( \begin{array}{c}
        U - i \lambda \left( h U + \Delta V \right) \\
        V + i \lambda \left( h^{*} V + \Delta^{*} U \right) 
            \end{array} \right).     \label{UVvariation}
\end{eqnarray}
Here, for the ease of the discussion, 
let us assume that the time increment $\Delta t$ is small enough 
so that we can identify the mean field Hamiltonian $h^{(n+1/2)}$ 
and mean pairing potential $\Delta^{(n+1/2)}$ in the TDHFB 
Hamiltonian ${\cal H}^{(n+1/2)}$ with $h$ and $\Delta$, respectively, 
at the time $t$.
Using the relations in Eq. (\ref{UVvariation}), the variations in the 
density $\rho$ in Eq. (\ref{rhoVV}) 
and pairing tensor $\kappa$ in Eq. (\ref{kapVU}) 
are represented, respectively, as
\begin{widetext}
\begin{eqnarray}
  \rho' &=& V'^{*} V'^{T} \nonumber \\
        &=& \left( V^{*} -i \lambda \left( \Delta U^{*} + h V^{*} \right)  \right)
            \left( V^{T} +i \lambda \left( U^{T} \Delta^{\dagger} + V^{T} h^{\dagger} \right)  \right) \nonumber \\
        &=& \rho -i \lambda [ h, \rho ] - i \lambda \left( - \Delta \kappa^{*} + \kappa \Delta^{*}  \right), 
                                                                                   \label{varrho}  \\
  \kappa' &=& V'^{*} U'^{T} \nonumber \\
        &=& \left( V^{*} -i \lambda \left( \Delta U^{*} + h V^{*} \right)  \right)
            \left( U^{T} -i \lambda \left( U^{T} h^{T} + V^{T} \Delta^{T} \right)  \right) \nonumber \\
        &=& \kappa -i \lambda \left( \Delta U^{*} U^{T} + h \kappa + \kappa h^{*} - \rho \Delta  \right).
\end{eqnarray}
\end{widetext}
Putting these expressions of the density and pairing tensor up to the 
first order in the parameter $\lambda$ into the expression of the energy in Eq. (\ref{Ehfb}),  
we have the variation of the energy as follows:
\begin{widetext}
\begin{eqnarray}
  \delta E &=& {\rm Tr} \bigl\{ -i \lambda h [h, \rho] 
          - i \lambda h \left( - \Delta \kappa^{*} + \kappa \Delta^{*} \right) \bigr\} \nonumber \\
        & & - \frac{i \lambda}{2} 
           {\rm Tr} \bigl\{ - \Delta^{*} \Delta - \Delta^{*} h \kappa - \Delta^{*} \kappa h^{*} 
                            + \Delta^{*} \Delta \rho^{*} + \Delta^{*} \rho \Delta \bigr\} \nonumber \\
        & & - \frac{i \lambda}{2} 
           {\rm Tr} \bigl\{   \quad \Delta^{*} \Delta + h^{*} \kappa^{*} \Delta + \kappa^{*} h \Delta 
                            - \Delta^{*} \rho \Delta - \rho^{*} \Delta^{*} \Delta \bigr\} \nonumber \\
        &=& 0, \label{e-conserv}
\end{eqnarray}
\end{widetext}
where the relations $h^{*} = h^{T}$, $\kappa^{T} = - \kappa$, 
and $\Delta^{T} = - \Delta$ are used. The notation Tr stands for 
taking the trace of the matrices.
From Eq. (\ref{e-conserv}), we see that 
we can integrate the TDHFB equation (\ref{hfbeq_UV}) with a conserved energy 
expressed as in Eq. (\ref{Ehfb}),
setting the time increment $\Delta t$ and intermediate 
TDHFB Hamilonian ${\cal H}^{(n+1/2)}$ adequately.
\par
When the Gogny interaction is used, 
there is a part in the mean potential $\Gamma$ 
which comes from the density-dependent term 
through the variation of the 
density matrix $\rho$, just as in Eq. (\ref{varrho}).
Owing to the parameter set of the Gogny interaction, 
there is no contribution of the density-dependent term to the 
pairing energy.
The contribution of the density-dependent term is 
included only in the mean-field Hamiltonian $h$ 
in the HFB Hamiltonian (\ref{hfbH}).
Then, the energy conservation relation (\ref{e-conserv}) 
holds when the Gogny interaction is adopted 
in the Hamiltonian (\ref{origH}).
\par
In the numerical calculation, we expand the exponential function in Eq. (\ref{tdhfb-solution}) 
in terms of the power series up to the tenth order in the time 
increment $\Delta t$. 
The intermediate TDHFB Hamiltonian ${\cal H}^{(n+1/2)}$ is 
made by using the predictor-corrector method at each time step:
In the predictor-corrector method, the predictor solutions 
$U'$ and $V'$ are calculated 
according to the method in Eq. (\ref{tdhfb-solution})
by using the TDHFB Hamiltonian ${\cal H}^{(n)}$ 
with the quantities $h$ and $\Delta$, which are 
made by using density $\rho^{(n)}$ 
and pairing tensor $\kappa^{(n)}$ at the time $t_n$, respectively. 
Then, using the predictor density $\rho' = V'^{*} V'^{T}$ 
and pairing tensor $\kappa' = V'^{*} U'^{T}$, the intermediate density 
$\rho^{(n+1/2)} = \left( \rho^{(n)} + \rho' \right)/2$ 
and the pairing tensor $\kappa^{(n+1/2)} = \left( \kappa^{(n)} + \kappa' \right)/2$ are 
made, which enter the intermediate TDHFB Hamiltonian ${\cal H}^{(n+1/2)}$. 
The corrector solutions $U''$ and $V''$ are calculated 
according to the method in Eq. (\ref{tdhfb-solution}) with the 
intermediate TDHFB Hamiltonian ${\cal H}^{(n+1/2)}$. 
From the ideal point of view, this process is repeated until 
the energy is conserved within a desired order. 
In the practical calculations, however, we stop the predictor-corrector 
iterations after the first two iterations to save the cpu time.

The initial condition we adopt in the present calculation is of the impulse type:
The static HFB solution $U_0$ and $V_0$ are changed into the initial matrices 
$U^{(0)}$ and $V^{(0)}$ by the relations given by
\begin{eqnarray}
  V^{(0)} &=& \exp \left( i \varepsilon {\bf Q} \right) V_0 
           = \sum_{\nu = 1}^{N_{max}} 
           \frac{i^\nu \varepsilon^\nu {\bf Q}^\nu}{\nu!} V_0,  \label{ini-1} \\
  U^{(0)} &=& \exp \left( -i \varepsilon {\bf Q}^{*} \right) U_0  
           = \sum_{\nu = 1}^{N_{max}} 
           \frac{i^\nu (-\varepsilon)^\nu {{\bf Q}^{*}}^\nu}{\nu!} U_0,  \nonumber \\
          & &                                                             \label{ini-2} 
\end{eqnarray}
where the expression ${\bf Q}$ stands for matrix representation 
of a multipole operator with respect to the numerical basis states. 
In the expression of the initial conditions (\ref{ini-1}) and (\ref{ini-2}), 
the exponential function is expanded into the power series 
with respect to the parameter $\varepsilon$ up to the $N_{max}$-th order.

\begin{figure}[htb]
\begin{center}
\includegraphics[width=7cm,angle=0]{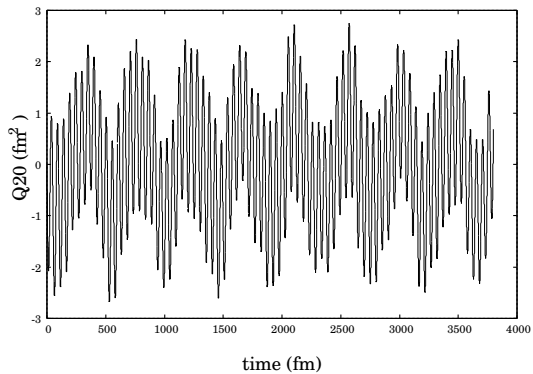}
\end{center}
\caption{Time dependence of expectation value of quadrupole moment. 
}
\label{quadosci}
\end{figure}

\begin{figure}[htb]
\begin{center}
\includegraphics[width=7cm,angle=0]{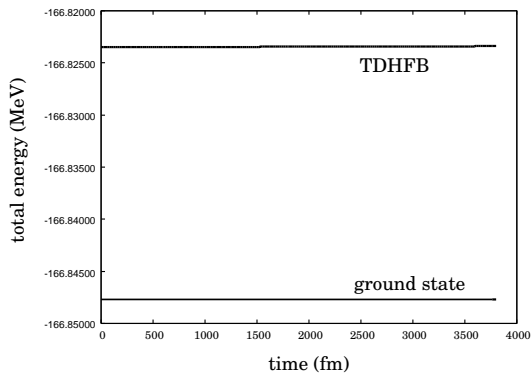}
\end{center}
\caption{Total energy vs time in quadrupole oscillation 
in Fig. \ref{quadosci}. Line labelled with "ground state" 
stands for the energy of the HFB ground state. 
Curve labelled with "TDHFB" is for the energy expectation value 
in the course of quadrupole oscillation.
}
\label{oscienergy}
\end{figure}

\begin{figure}[htb]
\begin{center}
\includegraphics[width=7cm,angle=0]{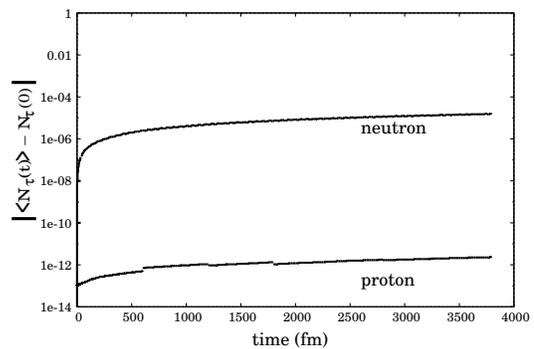}
\end{center}
\caption{Deviation of neutron (proton) number expectation value 
from accurate number 12 (8) in quadrupole oscillation in Fig. \ref{quadosci}.
}
\label{numb_consv}
\end{figure}

\begin{figure}[htb]
\begin{center}
\includegraphics[width=7cm,angle=0]{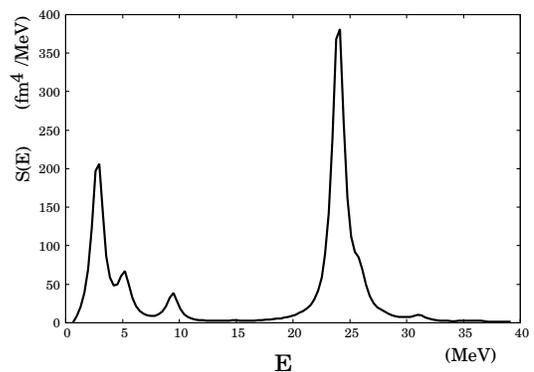}
\end{center}
\caption{Quadrupole strength distribution of quadrupole 
oscillation in Fig. \ref{quadosci}. Artificial width of 
0.6 MeV is used.
}
\label{strength}
\end{figure}

\section{Applications}
With the purpose of studying the feasibility of the TDHFB equation 
and the method of solution proposed in the previous section, 
we apply the method to the oxygen isotope ${}^{20}$O.
We adopt the Gogny force \cite{Gogny,Gogny2} with D1 parameter set 
in the two-particle interaction part 
in the Hamiltonian (1). 

As the numerical basis, we make use of the three-dimensional 
harmonic oscillator eigenstates. 
The two-particle matrix elements ${\cal V}_{\alpha \beta \gamma \delta}$ 
in the Hamiltonian (1) are calculated after the method 
which was used by Girod and Grammaticos \cite{Girod}. 
In the present calculations, we set the space of basis states so that 
the relation $n_x + n_y + n_z \leq N_{shell} = 4$ is satisfied, 
where $n_x$ ($n_y$, $n_z$) is the number of quanta 
of the harmonic oscillator basis states in the x (y, z) direction, respectively. 
The angular frequency parameters $\omega_x$, $\omega_y$, and $\omega_z$ 
of the harmonic oscillator basis states 
are optimized under the sphericity condition 
$\omega_x = \omega_y = \omega_z = \omega_0$ so that the HFB energy 
should be minimum.
We have set $\hbar \omega_0$ = 16.0 MeV.

In the initial conditions (\ref{ini-1}) and (\ref{ini-2}), 
we take the multipole operator ${\bf Q}$ to be 
a quadrupole operator $Q_{\alpha \beta} 
= \left( 2 z^2 - x^2 - y^2 \right)_{\alpha \beta}$, 
expressed as a matrix with respect to the basis states 
labelled by $\alpha$ and $\beta$.
The parameter $\varepsilon$ is put to be $1.0 \times 10^{-3}$, 
which is small enough so that the linearity of the oscillation 
with respect to $\varepsilon$ is satisfied.
The power series expansions of the exponential functions 
in Eqs. (\ref{ini-1}) and (\ref{ini-2}) are calculated up to 
the $N_{max}$-th order with $N_{max}$ put to be ten.

In the present calculations, we have omitted the Coulomb part 
in the two-body interactions, which leads to shorter cpu time.
For the moment, it might not be a draw-back to neglect the Coulomb part 
with the aim at studying the feasibility of the method under consideration.

In Fig. \ref{quadosci}, we display the time variation of the 
mass quadrupole moment $\langle 2 z^2 - x^2 - y^2 \rangle$ of ${}^{20}$O. 
The time increment $c \Delta t$ used in the 
calculation is 0.2 fm, and total time step is 19000. 
After the initial impulse, we can see regular small-amplitude oscillations 
take place. 

In Fig. \ref{oscienergy}, the total energy in the course of the 
oscillation is shown together with the ground state energy of the 
static HFB calculation.
The excitation energy 0.2 MeV is kept to a good extent in the 
integration process. 

In Fig. \ref{numb_consv}, the deviation of the expectation values 
of the nucleon number from the accurate values (8 protons and 12 neutrons) 
is displayed with respect to time. 
In ${}^{20}$O, the protons are in the normal state, whereas the 
neutrons are in the superconducting state. 
Therefore, the integration of the equations of motion 
of the proton orbitals is equivalent to the 
TDHF case, where the occupation probability of each one of the orbitals 
is exactly 1 or 0. 
Then, the total number of protons are conserved within 
$10^{-11}$. 
The neutron number in Fig. \ref{numb_consv} is kept 
up to around $10^{-5}$ in the present integration process.
This result illustrates that we can keep the unitarity of the time development 
operator in Eq. (\ref{tdhfb-solution}) within a practically 
satisfying order with the following set of 
parameters such as $N_{shell}$ = 4 and $c \Delta t$ = 0.2 fm in the 
case of the excitation energy 0.2 MeV.

In Fig. \ref{strength}, we display the strength function 
which is calculated by the Fourier transformation from the 
time series of the expectation value of the quadrupole  
operator in Fig. \ref{quadosci}. 
The locations of the main peaks in Fig. \ref{strength}  
are similar to the results in Ref. 6. 
The low-energy 2${}^{+}$ levels of ${}^{20}$O are located at  
1.7 MeV, 4 MeV, 5 MeV, and 10 MeV \cite{o20data}. 
The three low-energy peaks in Fig. \ref{strength} are 
expected to correspond to these levels.
Since the space of the basis states is not large enough, 
the energies corresponding to the peaks in Fig. \ref{strength} 
are somewhat higher than those of the observed levels.  

On the other hand, the ratio of the lowest energy peak 
around 3 MeV to the high energy one around 24 MeV is 
quite different from the result in Ref. 6. 
Considering that the effect of the continuum is included 
in the RPA calculations in Ref. 6, 
and our space in the present calculations 
is limited within the major quantum number $N_{shell}$ = 4, 
the differences of the low-energy peaks might come from the 
space of states taken in the calculations.

\section{Summary and Concluding remarks}
In this article, we have proposed a method to integrate 
the self-consistent TDHFB equation. 
We made use of an integration method 
which is widely used in the TDHF, i.e., a power series 
expansion of the time displacement operator at each time 
step. Adopting the Gogny interaction in the two-particle 
interaction parts in the Hamiltonian, we carried out the 
numerical calculations of the TDHFB equation in the case of 
oxygen isotope ${}^{20}$O. 

The accuracy of conservation of the excitation energy 
and the particle numbers is illustrated in the calculation of small-amplitude 
quadrupole oscillation, which is started from an impulse-type 
initial condition. 

The strength function of the quadrupole oscillation is calculated. 
The energies of the peaks of the strength function are similar to 
and somewhat higher than the experimental results of the low-energy 
2${}^{+}$ levels. 

The relative ratios of the peak heights, on the other hand, seem not to be enough
to discuss their physical contents, in comparison with the results given by the QRPA 
with continuum space included \cite{khan}.
One major reason of the situation might be the space of basis states 
in the present calculations, which is not large enough with the maximum major shell 
quantum number $N_{shell}$ = 4.
It is the foremost task to check the convergence of the 
results with respect to the cut-off $N_{shell}$ of the space of basis 
states. 

Since the basis states in the present calculations are 
the three-dimensional harmonic oscillator states, 
the method proposed in this article seems to be useful 
in describing the excitational modes in deformed nuclei.
The calculations in some deformed nuclei are now in progress. 
\begin{acknowledgments}
The authors thank Professor K.~Yabana and Dr. T.~Nakatsukasa for fruitful 
suggestions and comments, as well as encouragements. 
They are thankful to the members of the Nuclear Theory Group in the 
University of Tsukuba for daily discussions.
\end{acknowledgments}

\bibliography{Gognytdhfb-f}

\end{document}